\documentclass[twocolumn]{aastex631}

\shorttitle{EDGE-INFERNO}
\shortauthors{Andersson et al.}

\begin{document}

\title{EDGE-INFERNO: Simulating every observable star in faint dwarf galaxies\\ and their consequences for resolved-star photometric surveys}

\correspondingauthor{Andersson}
\email{eandersson@amnh.org}

\author[0000-0003-3479-4606]{Eric~P.~Andersson}
\affiliation{Department of Astrophysics,
American Museum of Natural History, 200 Central Park West, New York, NY 10024, USA}

\author[0000-0002-1515-995X]{Martin~P.~Rey}
\affiliation{Department of Physics,
University of Bath, Claverton Down, Bath, BA2 7AY, UK}
\affiliation{Sub-department of Astrophysics,
University of Oxford, Keble Road, Oxford OX1 3RH, United Kingdom}

\author[0000-0001-9546-3849]{Andrew~Pontzen}
\affiliation{Institute for Computational Cosmology, Department of Physics, Durham University, South Road, Durham, DH1 3LE, UK}

\author[0000-0003-2285-0332]{Corentin~Cadiou}
\affiliation{Lund Observatory, Division of Astrophysics, Department of Physics, Lund University, Box 43, SE-221 00 Lund, Sweden}

\author[0000-0002-4287-1088]{Oscar~Agertz}
\affiliation{Lund Observatory, Division of Astrophysics, Department of Physics, Lund University, Box 43, SE-221 00 Lund, Sweden}

\author[0000-0002-1164-9302]{Justin~I.~Read}
\affiliation{Department of Physics, University of Surrey, Guildford, GU2 7XH, UK}

\author[0000-0002-1349-202X]{Nicolas~F.~Martin}
\affiliation{Universit\'e de Strasbourg, CNRS, Observatoire astronomique de Strasbourg, UMR 7550, F-67000 Strasbourg, France}
\affiliation{Max-Planck-Institut f\"{u}r Astronomie, K\"{o}nigstuhl 17, D-69117 Heidelberg, Germany}

\begin{abstract}
Interpretation of data from faint dwarf galaxies is made challenging by observations limited to only the brightest stars. We present a major improvement to tackle this challenge by undertaking zoomed cosmological simulations that resolve the evolution of all individual stars more massive than $0.5\,{\rm M}_{\odot}$, thereby explicitly tracking all observable stars for the Hubble time. For the first time, we predict observable color-magnitude diagrams and the spatial distribution of $\approx 100,000$ stars within four faint ($M_{\star} \approx 10^5 \, \,{\rm M}_{\odot}$) dwarf galaxies directly from their cosmological initial conditions. In all cases, simulations predict complex light profiles with multiple components, implying that typical observational measures of structural parameters can make total V-band magnitudes appear up to 0.5 mag dimmer compared to estimates from simulations. Furthermore, when only small ($\lessapprox100$) numbers of stars are observable, shot noise from realizations of the color-magnitude diagram introduces uncertainties comparable to the population scatter in, e.g., total magnitude, half-light radius, and mean iron abundance measurements. Estimating these uncertainties with fully self-consistent mass growth, star formation and chemical enrichment histories paves the way for more robust interpretation of dwarf galaxy data.
\end{abstract}


\section{Introduction} \label{sec:intro}

The advent of deeper, wider photometric surveys in the last two decades (e.g., the Sloan Digital Sky Survey, and the Dark Energy Survey, hereafter, referred to as SDSS and DES) has revealed many, new, ever-fainter dwarf galaxies around the Milky Way (e.g., \citealt{Belokurov2007, Drlica-Wagner2015, Koposov2015}; see \citealt{Simon2019} for a review). 

These tiny galaxies ($M_{\star}\lessapprox10^{6}\,{\rm M}_{\odot}$) likely inhabit low-mass dark matter halos ($\approx10^{8-9}\,{\rm M}_{\odot}$; \citealt{Jethwa2018, Nadler2020}). Their mere existence provides sensitive constraints on dark matter models that suppress the growth of small-scale structures (e.g., warm dark matter, axions; \citealt{Nadler2019}). Furthermore, their low metallicities ($\rm [Fe/H]\approx -2$) provide a laboratory to study the first generations of stars and their chemical enrichment (\citealt{Frebel2015, Ji+2015}). Understanding how faint dwarf galaxies form and assemble in a $\Lambda$ cold dark matter ($\Lambda$CDM) universe is thus of broad importance and urgency, as the forthcoming Vera C. Rubin Observatory Legacy Survey of Space and Time (LSST) is expected to vastly expand the census of faint dwarf galaxies in the coming years (e.g. \citealt{Newton2018, Mutlu-Pakdil2021}).

In recent years, numerical models of low-mass galaxies have improved significantly \citep[see, e.g.,][]{Hu+2016, Wheeler2019, Agertz+2020, Smith2021}, even unlocking the ability to account for individual stars within a whole galaxy (e.g., \citealt{Hu+2017, Emerick+2018, Andersson+2020, Andersson+2021, Andersson+2023, Lahen+2020, Lahen+2023, Lahen+2024, Calura+2022, Gutcke+2021, Gutcke+2022a, Gutcke+2022b, Hislop+2022, Steinwandel+2023, Steinwandel+2024, Deng+2024}), as opposed to aggregating stars into a single stellar population particle. By explicitly drawing from the initial mass function (IMF) at stellar birth and modeling the stellar evolution tracks of each star, these models provide a much more detailed and robust account of how, when, and where feedback and metals are deposited into the interstellar medium (ISM). Furthermore, this novel model approach opens up the door for incorporating aspects of collisional dynamics present in star clusters that potentially affect both the ISM, galactic winds, and circumgalactic medium \citep[e.g., runaway stars and clustering of feedback, see][for a review]{Naab&Ostriker2017}. While high-order direct $N$-body integration is prohibitively computationally expensive for all stars in a galaxy over a Hubble time, its effects can be treated with sub-grid models \citep[e.g.,][]{Andersson+2020, Andersson+2023, Steinwandel+2023} using calibrations from highly detailed models applied in idealized set-ups \citep{Grudic+2021, Grudic+2022, Lahen+2024b, Polak+2024}.

However, the computational expense has constrained their applications to idealized, non-cosmological settings. This limits their ability to interpret redshift $z=0$ data as the cosmological mass growth history of a dwarf galaxy is key to shaping its observables (e.g. \citealt{Fitts2017,  Rey+2019b, Rey+2020, Rey+2022, Benitez-Llambay2021}). 

In the rare cases when individual-star models have included a full cosmological environment (e.g. \citealt{Gutcke+2022a, Gutcke+2022b, Calura+2022}), only massive (and thus less numerous) stars ($\gtrapprox 4 \, \,{\rm M}_{\odot}$) are individually sampled to ease the computational load. But slow-moving stellar mass loss from low-mass evolved stars has characteristic velocities ($\approx 10 \, \rm km \, s^{-1}$;  \citealt{Hofner&Olofsson2018}) comparable to the depth of gravitational potential wells of faint dwarf galaxies. As a result, winds from e.g. asymptotic giant branch stars (AGBs) likely play a role in chemical enrichment and possibly delay the restart of star formation in galaxies quenched by reionization (e.g. \citealt{Rey+2020}). Furthermore, the lowest mass galaxies are star formation quenched during the epoch of reionization and hence their mass budget in massive stars by redshift zero is fully depleted. Instead dwarf galaxy properties are characterized through the imaging and spectroscopy of the light from their old, evolved low-mass stars, motivating a complete account of the stellar population.

In this paper, we combine all these advances for the first time to take a major step toward interpreting resolved-star data from dwarf galaxies. We perform high-resolution cosmological simulations within the EDGE\footnote{Engineering Dwarfs at Galaxy formation's Edge} framework, combining them with the star-by-star INFERNO\footnote{INdividual stars with Feedback, Enrichment, and Realistic \\ Natal mOtions} model \citep{Andersson+2023}. With this, we achieve a high numerical resolution ($3\,{\rm pc}$ at $z=0$) that allows us to capture the evolution of supernova (SN) explosions self-consistently over the full cosmological formation of the dwarf, while individually sampling all stars down to $0.5 \, \,{\rm M}_{\odot}$ (main sequence lifetime $>14\, \rm Gyr$). This ensures that all major sources of stellar feedback are accounted for on a star by star basis, while self-consistently populating the observable spatial and color-magnitude distribution at $z=0$, at which point stars born before reionization and with mass $\sim0.8\,{\rm M}_{\odot}$ evolve to the giant branch (thus become bright enough to be observable).

We describe these simulations and methods in Sections~\ref{sec:edge-inferno} and \ref{sec:stellar-mass-growth}  and, in Section~\ref{sec:observing-simulations}, present the observational implications of directly predicting resolved-star observables from cosmological initial conditions.

\section{Star-by-star cosmological simulations}\label{sec:edge-inferno}

\begin{figure}
    \centering
    \includegraphics[width=\linewidth]{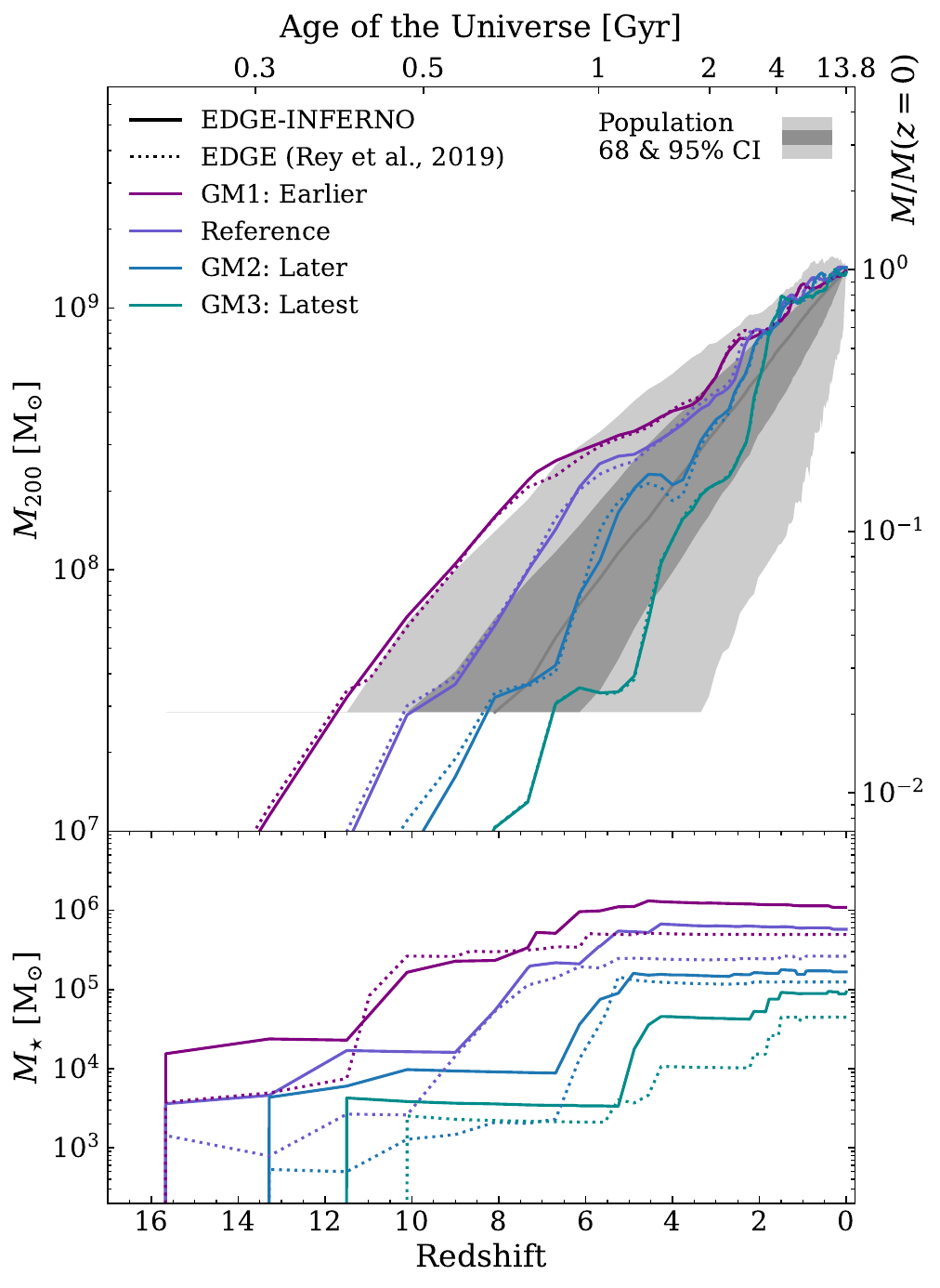}
    \caption{Halo mass ($M_{200}$, top panel) and stellar mass (bottom panel) as a function of redshift (bottom axis) and age of the Universe (top axis) showing the four galaxies simulated star-by-star (filled lines) and with star particles \citep[$300\,{\rm M}_{\odot}$ resolution, dotted lines][]{Rey+2019b}. Dark (light) gray-shaded region encloses $95\%$ ($68\%$) of halo growth histories in the entire simulation box ($50\,{\rm Mpc}^3$) from which our initial conditions were selected. The right y-axis of the top panel shows the fraction of mass relative to the halo mass at redshift zero.} 
    \label{fig:mass_growth}
\end{figure}

We perform zoomed, cosmological simulations of a low-mass dark matter halo ($M_{200}=1.5\times10^{9}\,{\rm M}_{\odot}$ at $z=0$). Using the genetic modification technique (\citealt{Roth+2016, Rey&Pontzen2018, Stopyra+2021}), we modify the initial conditions of this halo to engineer four different early growth histories for this object, while maintaining the same halo mass at $z=0$ and large-scale cosmological environment \citep[see][for further details of the genetically-modified initial conditions]{Rey+2019b}. 

Figure~\ref{fig:mass_growth} shows the halo mass growth history (top panel) from our new suite of simulations (filled) matching the original EDGE simulations (dotted \citealt{Rey+2019b}). We define halo mass as $M_{200}$, defined as the mass enclosed with the radius $r_{200}$ where the density is 200 times the critical density of the Universe. Genetic modifications of this halo are designed to scan through the 68-95\% population of early mass-growth histories for dark matter halos of this mass (gray contours; see details of the calculation in \citealt{Rey+2019b}), at fixed halo mass today. Compared to randomly sampling halos, this genetically-modified setup allows us to efficiently probe the scatter in observable properties while fixing the main control variable ($M_{200}$).

For this work, we re-evolve these initial conditions to $z=0$ using the adaptive mesh refinement code \textsc{ramses} \citet{Teyssier2002} and the star-by-star model \textsc{inferno} described in depth in \citet{Andersson+2023}. The hydrodynamical, gravity, and thermochemistry setup are the same as in  \citet{Andersson+2023}, while the refinement strategy is identical to \citet{Rey+2019b, Agertz+2020}. The median resolution is $3\,{\rm pc}$ within the ISM of the galaxy (dark matter particle mass of $980\,{\rm M}_{\odot}$).

Star formation follows \citet{Agertz+2020}, assuming stars are born in quantities of $500\,{\rm M}_{\odot}$ in cells with gas density $>300\,{\rm cm}^{-3}$ and temperature $<10^3\,{\rm K}$. Spawning stars in groups of $500\,{\rm M}_{\odot}$ minimizes issues related to incomplete sampling of the IMF \citep[][]{Smith2021}. In addition to the gas velocity, stars are stirred with a velocity dispersion of $0.01\,{\rm km\,s}^{-1}$ at birth \citep{Andersson+2023}. This velocity dispersion is below the resolution of the gravitational softening (cell size) and serves as a numerical approximation to allow particles with identical initial position end up on trajectories that eventually diverge due to small but repeated dynamical perturbation.

The key novelty of this work is the introduction of individual star particles once a star formation has been flagged. We draw masses from a \cite{Kroupa2001} IMF using the sampling technique from \cite{Sormani+2017}. Critically, all stars with masses $\geq 0.5\,{\rm M}_{\odot}$ are explicitly sampled and inject energy, momentum, mass, C, N, O, Mg, Al, Si, and Fe. Stars with masses $\le 0.5\,{\rm M}_{\odot}$ are expected to remain on their main sequence for the Hubble time and are thus aggregated into a low-mass, passive particle. Although important, stellar multiples (e.g., binaries) are not accounted for in our model (see further discussion in \citealt{Andersson+2023}).

Stellar feedback in our models includes SNe (core-collapse and Ia) and stellar winds (from OB type and AGB stars) implemented as in \cite{Andersson+2023}. While winds are a sub-dominant component of the feedback budget \citep[in particular at low metallicity, see, e.g.,][]{Vink+2001, Mokeim+2007, Vink2011}, they are an important contributor to metal enrichment and can provides a prompt source of energy injection affecting local star formation \citep{Andersson+2024}. Radiation is not included in our suite of simulations (or in the simulations from \citealt{Rey+2019b} we compare to) and is expected to further regulate star formation, systematically decreasing the total stellar masses of our galaxies (\citealt{Agertz+2020}, Rey et al. in prep). We read the chemical yields and mass-loss of each star as a function of mass and metallicity using tabulated values from \textsc{NuGrid} \citep[][]{Pignatari+2016, Ritter+2018} for core-collapse SNe and stellar winds, and from \citet{Seitenzahl+2013} for SNe Ia. In all cases, we calculate the main sequence lifetime of a star following the fitting function from \cite{Raiteri+1996}. SNe Ia follow the time-delay distribution for field galaxies from \cite{Maoz&Graur2017} assuming a rate of $I_{\rm Ia}=2.6\times10^{-13}\,{\rm yr}^{-1}\,{\rm M}_{\odot}^{-1}$ starting $38\,{\rm Myr}$s after birth. 

As in \cite{Andersson+2023}, massive stars ($8 \leq m_{\star} \leq 60 \,{\rm M}_{\odot}$) inject fast ($1000\,{\rm km\,s}^{-1}$) winds while on the main sequence and end their life with core-collapse directly injecting material and $10^{51}\,{\rm erg}$ if $m_{\star} \leq 30\,{\rm M}_{\odot}$ (assuming direct collapse to black hole otherwise). While the high resolution of our simulations allows direct injection of SNe energy to obtain accurate capture of the final momentum \citep{Ohlin+2019} in almost all cases ($>96\%$), a switch to terminal momentum injection \citep{Kim&Ostriker2015} is applied in the rare cases when the resolution is not sufficient \citep[see][for further details]{Agertz+2020, Andersson+2023}.

Low mass stars ($<8\,{\rm M}_{\odot}$) eject slow ($10\,{\rm km\, s}^{-1}$) winds with a constant mass loss rate \citep[$10^{-5}\,{\rm M}_{\odot}\,{\rm yr}^{-1}$][]{Eriksson+2014, Hofner&Olofsson2018} during their AGB phase. Wind injection stops when their stellar mass follows the initial-to-final mass relation from \cite{Cummings+2016}.

\section{Stellar mass growth}\label{sec:stellar-mass-growth}

The lower panel of Figure~\ref{fig:mass_growth} shows the stellar mass growth of our dwarf galaxies. With $M_{\star}(z=0) = 10^5 - 10^6 \,{\rm M}_{\odot}$ and $M_{200} = 1.5 \times 10^9 \, \,{\rm M}_{\odot}$, our galaxies are within the 2$\sigma$ scatter inferred for local Milky Way satellites (e.g. \citealt{Nadler2020}, Figure 6).  

Compared with the traditional stellar-particle approach (dotted; \citealt{Rey+2019b}), swapping to individual-star feedback systematically increases $M_{\star}$ at all redshifts, by at most a factor of 2. Such variations are well within theoretical uncertainties at this mass scale (see, e.g., the discussion in \citealt{Agertz+2020}, and review in \citealt{Sales+2022}) which should be expected given the extensive updates to the chemical enrichment channels (and thus gas cooling) and updates to the stellar feedback implementation.\footnote{Even if the total feedback budget and injection mechanisms are fixed, the location and clustering of individual stars can be significantly different and affect galactic outflows (e.g. \citealt{Andersson+2020, Andersson+2023, Smith2021}).} 

Whether using individual stars or stellar particles, earlier-forming dwarfs systematically assemble more stellar mass by $z=0$. This confirms that relative effects between mass growth histories are robust to model changes, even if astrophysical assumptions leave residual uncertainties in the absolute normalization of the stellar mass (see also discussion in \citealt{Rey+2019b}). Combined, these consistency checks allow us to confidently turn to harnessing the new explicit connection between assembly and resolved-stars observables.

\section{Direct predictions of dwarf galaxies' resolved star observables}\label{sec:observing-simulations}

\begin{figure*}
    \centering
    \includegraphics[width=\linewidth]{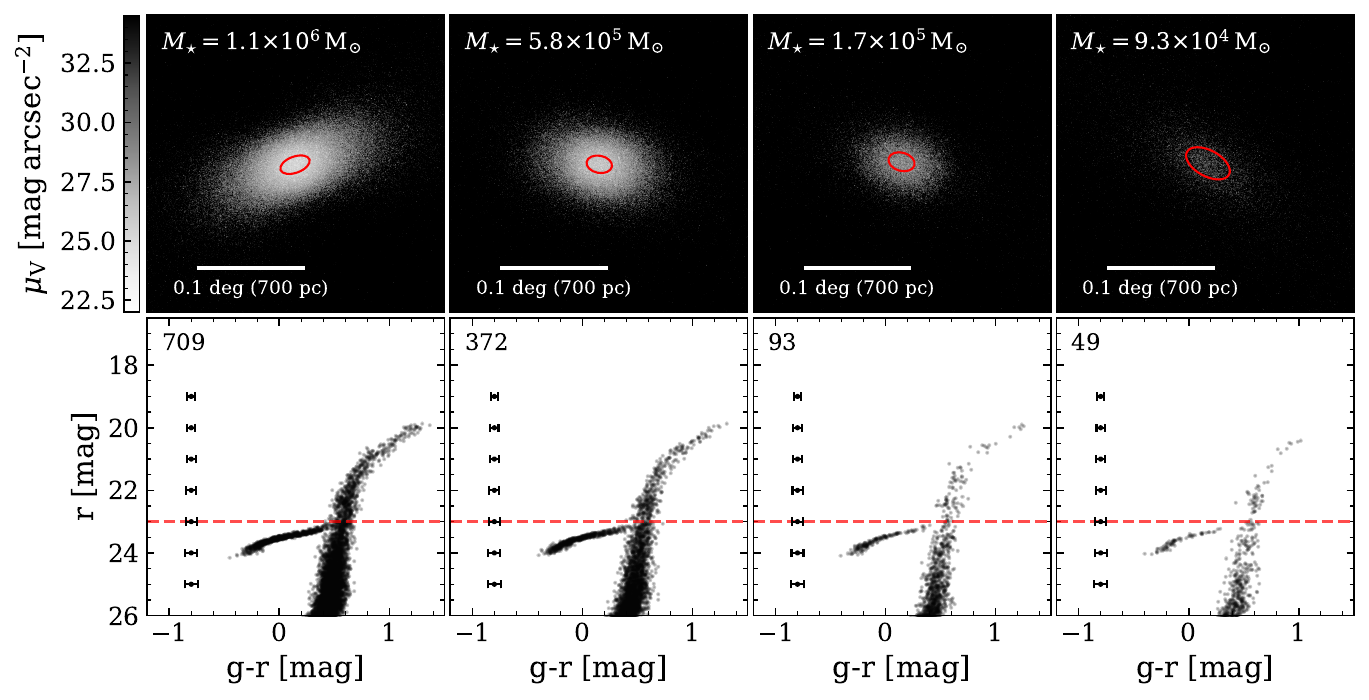}
    \caption{V-band surface brightness maps for each galaxy (top row) at $400\,{\rm kpc}$ distance at a resolution of $0.3"$ per pixel (equivalent to DES). Compared with traditional stellar population approaches, modeling every star with $\geq 0.5 \, \,{\rm M}_{\odot}$ greatly increases stellar sampling throughout the galaxy, both in the center and the outskirts. It also directly predicts the simulated dwarf's color-magnitude diagram (here scaled by a $400\,{\rm kpc}$ distance modulus in r and g band magnitudes, error bars reflecting the observationally motivated applied scatter) and spatial distribution, enabling the estimation of structural properties as an observer would (red ellipses show the elliptical half-light radius; see text for details). As galaxies become less massive (left to right), fewer stars stay above a given magnitude limit (e.g. red-dashed line for a DES-like survey), in turn introducing greater scatter in their derived properties (Figure~\ref{fig:rhalf_FeH_MV}).}
    \label{fig:surface_brightness}
\end{figure*}

\subsection{Color-magnitude diagrams}

Figure~\ref{fig:surface_brightness} highlights the vastly improved stellar sampling of our simulated galaxies, showing the surface brightness (top row) as if the galaxies are observed at $400\, {\rm kpc}$ with the resolution of DES. Each star is assigned stellar properties (e.g., luminosity) derived by interpolation of \textsc{parsec} \citep{Bressan+2012} isochrones (see Appendix~\ref{sec:isochrones_LF} for details) accounting for stellar mass, age, and metallicity. In the top panels of Figure~\ref{fig:surface_brightness}, low-mass stellar aggregates are accounted for by sampling the IMF in the range $0.08$ and $0.5\,{\rm M}_{\odot}$. Bright individual stars are visible at this distance. 

The bottom row of Figure~\ref{fig:surface_brightness} shows each galaxy's respective color-magnitude diagram (CMD) for r and g band magnitudes scaled with a $D = 400\,{\rm kpc}$ distance modulus. For each CMD, we add a scatter similar to the typical noise associated with the equivalent observation (left points in each panel, \citealt{Homma+2018}). The dashed red line visualizes the typical point-source limit accessible with the DES \citep[][]{Simon2019}. As the stellar mass decreases (left to right), the number of stars above this observable threshold decreases (top-left number). Such observable constructs are a natural byproduct of our simulations, as is predicting the spatial location and chemistry of the brightest, observable giant stars. This enables us to derive dwarf galaxy properties directly with observational tools, to which we turn now. We specifically chose $D = 400\,{\rm kpc}$ to match the tens to hundreds of stars typically observed in dwarfs with a DES-like setup. We vary these assumptions in Appendix~\ref{sec:SDSS_VRT_comparison}.

All our dwarf galaxies have diffuse light profiles, with hierarchical merging leading to distant bound members up to $\geq 5 \, r_h$ similar to those reported in recent observations (e.g., \citealt{Chiti2021, Sestito2023}). As we will see, these multi-component spatial distributions can bias the recovery of faint dwarf structural parameters, particularly when few stars are bright enough to be observed. Such outer stellar halo for faint dwarf galaxies was already pointed out in simulations with stellar particles (e.g. \citealt{Tarumi+2021, Goater+2024}). In theory, post-processing techniques sampling the IMF could be used to transform such stellar particles into individually resolved stars (e.g. \citealt{Krumholz+2015, Grand+2018, Sanderson+2020, Gray+2024}) and test observational pipelines. However, up-sampling techniques can create strong phase-space artifacts, especially when stellar sampling is limited (see e.g. discussion in \citealt{Lim+2024}). The vastly improved stellar sampling of our resolved-star simulations -- $\approx 100,000$ stars per galaxy compared to $\approx 500$ stellar particles in \citep{Rey+2019b} -- significantly improves the robustness in predicting the kinematic and chemical properties stellar distribution. We turn next to quantifying how this impacts the observational recovery of structural parameters and will explore radial trends and properties of distant members in future work.

\subsection{Structural properties}

\begin{figure}
    \centering
    \includegraphics[width=\linewidth]{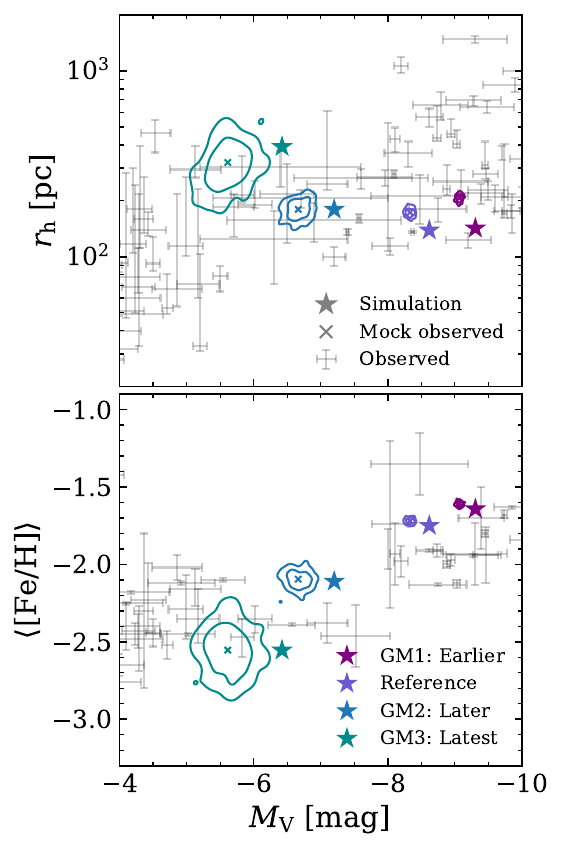}
    \caption{Half-light radius (top panel) and average iron abundance (bottom panel) as a function of total absolute V-band magnitude of our simulations, compared to Local group data (gray). Obtaining $r_h$ and $M_{\rm V}$ via mock observation of resolved stars (crosses) provides estimates of total V-band magnitude that are systematically lower than the magnitude derived from the simulations (star symbols). This apparent dimming arise from light profiles emerging from cosmological formation histories that are more complex than typically assumed for observed dwarfs. Stochastically re-sampling the CMD for the same history (contours around crosses) introduces a noise term, as different realizations of the light and metallicity distribution get over the magnitude limit. This noise becomes comparable to the population scatter for our fainter objects, with only a few tens of observable stars. Observed data from the compilations of \citet{McConnachie2012, Kirby+2013, Kirby2014, Simon2019} augmented with individual candidates and detections from \citet{Torrealba2016, Torrealba2018, Torrealba2019, Homma+2018, Homma2024, Mau2020, Bennet2022, Richstein2022, Sand2022, Cerny2023DELVE, Cerny2023DELVE6,Cerny2023PegIV, Jones2023, Jones2024, McQuinn2023, Collins2024, Li2024, MartinezDelgado2024, McNanna2024, Smith2023, Smith2024Faintest, Tan2024}.
    }
    \label{fig:rhalf_FeH_MV}
\end{figure}

We derive the half-light radius, total magnitude, and average iron metallicity ($r_h$, $M_{\rm V}$, $\langle[{\rm Fe/H}]\rangle$, respectively) as an observer would. Using the projected positions ($\alpha_i,\beta_i$) for each star $i$ brighter than the assumed limit (r=23 mag), we fit exponential profiles to the projected stellar number density (see \citealt{Martin+2008, Martin+2016} and Appendix~\ref{sec:LuminosityAndStructure} for further details). Red ellipses in Figure~\ref{fig:surface_brightness} show the best-fit elliptical half-light radius $r_h$ for each galaxy, with their stellar density profiles and best-fit parameters shown in Figure~\ref{fig:profile_23mag400kpc}. $\langle[{\rm Fe/H}]\rangle$ is the average of stars above the magnitude limit. 

Figure~\ref{fig:rhalf_FeH_MV} shows the derived $M_{\rm V}$, $r_h$ and $\langle[{\rm Fe/H}]\rangle$ with these procedures (colored crosses) compared to observational data. We also compare to `simulated' values (colored stars), determined by counting all stars within $r_{200}$ for $M_{\rm V}$ and $[{\rm Fe/H}]$, and interpolating the spherical half-light radius for $r_h$. These `simulated' definitions are intrinsically different from the observational quantities, but are commonly used across the literature to compare models to data (e.g. \citealt{Wheeler2019, Agertz+2020, Rey+2020}) and provide a complete account of each galaxy's stars. For all galaxies, the total magnitudes are under-estimated using the observational approach, by up to $0.3$ mag for our brightest objects. This should be expected because a magnitude-limited observation will miss stars available to the simulation. We emphasize that the lower magnitude is not due to missing the light from stars below the observed limit, which is accounted for by sampling luminosity functions (see Appendix~\ref{sec:LuminosityAndStructure}). Instead, the missing light is from stars not accounted for by the fitted exponential profile. Furthermore, brighter dwarfs (with more observed stars) remain biased in $M_{\rm V}$ and also show less accurate $r_h$ determinations (up to $\approx1.5 \times$ offset).

All our galaxies exhibit surface density profiles with multiple components including a roughly exponential center and an additional extended halo (see Figure~\ref{fig:profile_nobg} and Appendix~\ref{sec:LuminosityAndStructure} for details). When fitted to a single exponential, the extended population is not accounted for, biasing $M_{\rm V}$ to fainter values and $r_h$ to more compact values. $\Lambda$CDM structure formation predicts that even faint dwarf galaxies exhibit extended halos and distant members (Figure~\ref{fig:surface_brightness}, see also, e.g., \citealt{Tarumi+2021, Goater+2024}), so this inadequacy should be expected. Furthermore, the physical extended halo of the dwarf galaxy is largely indistinguishable from the typical background assumed for Milky Way observations ($\Sigma \approx 10^{-1} \, \text{arcmin}^{-2}$; see Equation~\ref{eq:exp_profile} and \citealt{Martin+2008, Martin+2016}). This gives the appearance of a good fit of the stellar population on the sky, but misses physical members of the dwarf galaxy (see Figure~\ref{fig:profile_nobg}) resulting in further systematic off-sets in $M_{\rm V}$ compared to the values estimated directly from the simulations. Distinguishing background stars from members will remain a challenge for real observations in the Local Group, but our simulations provide a first estimate of the biases when comparing models and observations. Furthermore, our simulated dwarfs evolve in the field and are therefore less prone to tidal stripping compared with the satellite galaxies observed in the Local Group. We leave testing how our galaxies evolve in an external potential field to future work.

Another source of uncertainty when estimating dwarf galaxy properties comes from the inherent stochasticity associated with which stars reside above the magnitude limit at $z=0$, and are thus bright enough to be observed (CMD shot noise). Observed stars only provide a subset of the true spatial distribution, light profile, and metallicity distribution of the whole dwarf galaxy. 

The exact CMD realization will depend on the initial mass of each low-mass star, its precise metallicity, and age. All these quantities are deterministic in our simulations, providing us with one example realization of the CMD for a given dwarf galaxy's history.

Nonetheless, we can estimate the magnitude of this noise term by perturbing the initial masses of each of our stars. In other words, the true spatial, age, and metallicity distributions are unchanged, but we resample which stars ascend the giant branch. We achieve this by resampling the IMF 100 times, in very fine bins ($0.25\,{\rm M}_{\odot}$) to minimize the impact that a different mass has on the dynamical evolution of any given star throughout the simulation. We then re-fit all properties as previously and show the 68-95$\%$ confidence intervals as contours in Figure~\ref{fig:rhalf_FeH_MV}. 

This noise has limited impact on bright galaxies with a well-sampled CMD (purple and violet), where it is far outweighed by the inaccurate $r_h$ and $M_{\rm V}$ from fitting. For fainter galaxies (cyan and blue) with only a few tens of visible stars, this shot noise introduce significant uncertainty, with $2\sigma$ shifts of up to $0.4\,{\rm mag}$, $230\,{\rm pc}$, $0.3\,{\rm dex}$ for $M_{\rm V}$, $r_h$ and $\langle[{\rm Fe/H}]\rangle$, respectively. 

These numbers are highly significant to the interpretation of dwarf galaxy data -- they are comparable to the population scatter in the $\langle[{\rm Fe/H}]\rangle$ - $M_{\rm V}$ plane when galaxies only have a few tens of observable stars. The number of detected member stars, and thus the extent of this noise, will depend on the specifics of each observation. But, even with modern facilities, faint dwarf galaxies often have $<50$ candidate member stars (e.g. \citealt{Cerny2023DELVE, Homma2024, Smith2024}), hinting that some of the observed scatter could be explained through CMD shot noise. Furthermore, this effect will be compounded by the need for spectroscopic follow-up. Here, all observable stars above a magnitude limit are considered when estimating a mean, while in reality, only a subset of stars can be followed up with high-resolution spectroscopy to determine $[{\rm Fe/H}]$ and metal abundances. We emphasize that when the metallicity distribution is sampled by a few tens of stars, a mean value for, e.g., iron abundance is a poor constraint on models. We provide a detailed analysis of specific abundance features, their observability, and the uncertainty of varying element yields in Andersson et al. (in prep.).

\section{Discussion and Conclusion}

We present a novel set of cosmological simulations of small dwarf galaxies with unprecedented detail, tracking and evolving all stars observable throughout the Hubble time on a star-by-star basis. With this, we self-consistently predict how $z=0$ resolved-star observables (e.g. observable spatial distribution, and CMD; Figure~\ref{fig:surface_brightness}) arise from the cosmological mass growth, star formation and chemical enrichment history of each dwarf galaxy. This allows us to directly apply observational parameter estimation to test the accuracy and precision of these pipelines on realistic objects (Figure~\ref{fig:rhalf_FeH_MV}).

In all four of dwarfs, hierarchical assembly creates light profiles with at least two different components. This leads to systematic offsets when recovering structural parameters with methods assuming a single exponential for profiles (e.g \citealt{Martin+2016}), with the total V-band magnitude being shifted by up to $0.5\,{\rm mag}$ compared to the value derived from simulations with commonly used methods. This outer component is faint, indistinguishable from background, and likely affected by tidal stripping in galaxies that reside in group environments. In future work, we will investigate how this can be accounted for and if methods including different density models \citep[e.g.,][]{Plummer1911, Sersic1963, King1966} including multiple components can give better estimates of the light profiles for observed dwarfs.

In realistic cases of observed faint dwarf galaxies, few member stars are bright enough to be confidently assigned to the galaxy. This introduces an intrinsic uncertainty to recovering the dwarf's galaxy underlying light profile and metallicity distribution from observationally-estimated properties. We show that when $\leq 100$ members stars are observable, CMD shot noise introduce significant uncertainty. This uncertainty makes interpretation particularly challenging in the $\langle[{\rm Fe/H}]\rangle$ - $M_{\rm V}$ plane, where it is comparable to the population scatter.

This work marks a significant step towards bridging simulations and observations of dwarf galaxies. The differences in quantities estimated using techniques commonly applied to survey pipelines compared to those commonly applied to simulations uncovered in this work have direct implications on how we map observables such as $M_{\rm V}$ (and thus $M_{\star}$) and $\langle[{\rm Fe/H}]\rangle$ onto cosmological dark matter halos, and in turn constrain the nature of dark matter \citep[e.g.,][]{Nadler2019} or the strength of galactic outflows \citep[e.g.,][]{Agertz+2020} with faint dwarf galaxies. While the magnitude of these differences is left to be quantified across a wide range of surveys for galaxies at all distances, this work underscores the importance of self-consistently accounting for the full cosmological history rather than assuming idealized objects when testing observational pipelines \citep[but see, e.g.,][]{Mutlu-Pakdil2021}. The advent of individual-star models such as \textsc{inferno}, and their application to cosmological simulations of galaxies down to low-mass stars, will enable these quantification's in the near-future and unlock much more robust interpretation of forthcoming data from e.g. the LSST.

\begin{acknowledgments}

We thank the anonymous referee for comments and suggestions that greatly improved the quality of this work. We acknowledge Mordecai-Mark Mac Low for helpful comments improving the quality of this manuscript.

EA acknowledges support from NASA ATP grant 80NSSC24K0935 and NSF grant AST23-07950. MR is supported by the Beecroft Fellowship funded by Adrian Beecroft. OA acknowledges support from the Knut and Alice Wallenberg Foundation, the Swedish Research Council (grant 2019-04659), and the Swedish National Space Agency (SNSA Dnr 2023-00164). AP was supported by the European Research Council (ERC) under the European Union’s Horizon 2020 research and innovation programme (grant agreement No. 818085 GMGalaxies). JIR would like to acknowledge support from STFC grants ST/Y002865/1 and ST/Y002857/1.

This work used the DiRAC Data Intensive service (DIaL2 / DIaL) at the University of Leicester, managed by the University of Leicester Research Computing Service on behalf of the STFC DiRAC HPC Facility (www.dirac.ac.uk). The DiRAC service at Leicester was funded by BEIS, UKRI and STFC capital funding and STFC operations grants. DiRAC is part of the UKRI Digital Research Infrastructure.

We thank the Astrophysics Data Service funded by NASA under Cooperative Agreement 80NSSC21M00561 and the arXiv preprint repository for providing services that were used extensively in this work.

\end{acknowledgments}

\software{\textsc{genetIC} \citep{Stopyra+2021},
          \textsc{ramses} \citep{Teyssier2002},
          \textsc{pynbody} \citep{Pontzen2013, Pontzen2022},  
          \textsc{tangos} \citep{Pontzen2018}, 
          \textsc{numpy} \citep{vanderWalt2011, Harris2020},
          \textsc{scipy} \citep{Virtanen2020},
          \textsc{emcee} \citep{Foreman-Mackey+2013},
          \textsc{matplotlib} \citep{Hunter2007},
          \textsc{jupyter} \citep{RaganKelley2014}
          }

\appendix

\section{Isochrones and luminosity functions}\label{sec:isochrones_LF}

We use isochrones and luminosity function derived from the \textsc{parsec} web-interface\footnote{\hyperlink{http://stev.oapd.inaf.it/cmd}{http://stev.oapd.inaf.it/cmd}} applying \textsc{parsec} version 1.2S \citep{Chen+2014, Chen+2015, Tang+2014} with \textsc{colibri} version S\_37 for TP-AGB end-phase evolution \citep{Marigo+2013, Rosenfield+2016, Pastorelli+2019, Pastorelli+2020}. Circumstellar dust is included using scaling relations from \citet{Marigo+2008} with $60\%$ silicate and $40\%$ AlO for M stars and $85\%$ AMC and $15\%$ SiC for C stars \citep{Groenewegen2006}. Long-period variability during RGB and AGB is included following \citet{Trabucchi+2021}. For magnitudes, we used photometric systems \textit{UBVRIJHK} \citep{Bessell1990, MaizApellaniz2006} and \textit{ugriz} from SDSS. We do not consider dust extinction for any of our results.

\section{Method for determining galaxy luminosity and structural properties}\label{sec:LuminosityAndStructure}

We determine the structural properties of the galaxies by the technique presented in \citeauthor{Martin+2008} (\citeyear{Martin+2008}, see also \citealt{Martin+2016}). Given a distance modulus and magnitude limit, the observable stars $i$ in each galaxy are projected to sky coordinates ($\alpha_i,\beta_i$). An additional set of stars is added to this group by uniformly sampling a background stellar field around the galaxy. For all examples presented throughout this work we use a background of $0.1$ star per arcmin$^2$. While the number density of the background is somewhat arbitrary, it is of the order expected in the Local Group and is expected by the fitting model. As in \citet{Martin+2008}, the projected number density in units star per arcmin$^{-2}$ given by 
\begin{equation}\label{eq:exp_profile}
    n(r) = \underbrace{\frac{1.68^2N_{\star}}{2\pi(1-e)r_h^2} \exp(-1.68r/r_h)}_{\rm galaxy} + \underbrace{(N_{\rm tot} - N_{\star}) / (3600\pi)}_{\rm background},
\end{equation}
where $N_{\star}$ is the number of stars in the galaxy, $e$ is eccentricity, $r_h$ is the half-density radius in units arcmin, $N_{\rm tot}$ is the total number of stars in the sample, and $r$ is the elliptical radius centered on the galaxy coordinates ($\alpha_0,\beta_0$) rotated by a north-to-east angle $\theta$ such that the major axis aligns with the position axis. We define elliptical radius as in \citet{Martin+2008}, using ellipticity as one minus the ratio between scale length along the minor and the major axis. The fitting procedure use applies the Markov Chain Monte Carlo sampler \citep{Goodman&Weare2010} implemented in \textsc{emcee} \citep[][]{Foreman-Mackey+2013} using 64 walkers and 5000 steps to determine the maximum of the log-likelihood function $\mathcal{L}(\alpha_0, \beta_0, \theta, e, r_h, N_{\star}) = \prod_i n(r_i)$. 

To determine total magnitude, we use luminosity functions (see Appendix~\ref{sec:isochrones_LF}) scaled to the assumed distance and sample stars from an IMF \citep{Kroupa2001} until the number of stars with a magnitude brighter than the set limit reaches the best fit for $N_{\star}$. In this case, the luminosity function is derived assuming the age and metallicity to those of the mean of the galaxy. This provides a complete stellar sample used to calculate the total magnitude in a given band. To ensure robustness of these estimates we perform this exercise 1000 times for each galaxy and compute the mean and standard deviation. We note that in all cases, this standard deviation is negligible compared to other errors.

\section{Estimating structure parameters of simulated data}\label{sec:SDSS_VRT_comparison}

\begin{figure*}
    \centering
    \includegraphics[width=\linewidth]{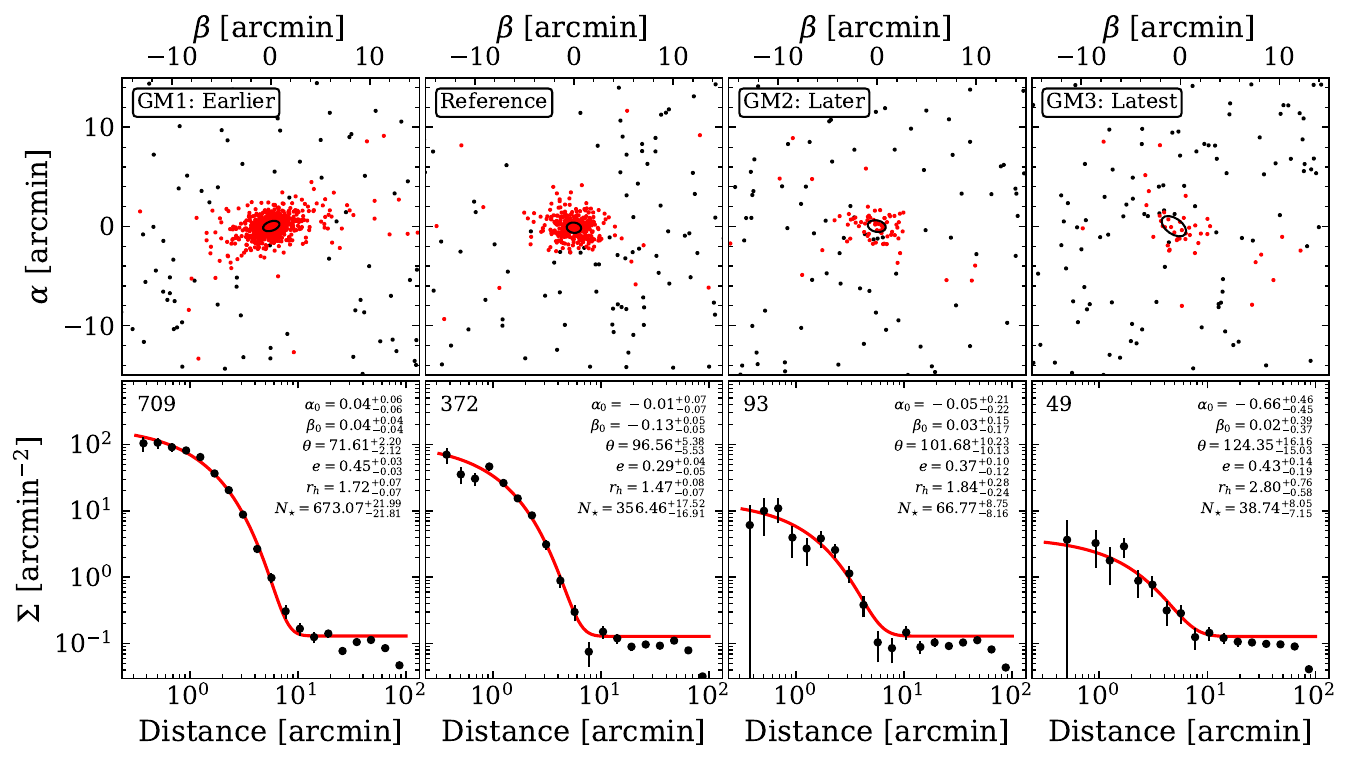}
    \caption{The projected surface number density of stars brighter than 23 magnitudes in r-band at $400\,{\rm kpc}$ as a function of projected distance from galaxy center (black points with error bars indicating Poisson noise). Red lines show profiles (Equation~\ref{eq:exp_profile}) fitted by a maximum likelihood search using \citep[\textsc{emcee}][]{Foreman-Mackey+2013} with parameters and uncertainty from the fitting procedure written for each galaxy in each panel. The $\alpha_0,\beta_0$ are the offsets from the halo center (determined through a shrinking sphere applied to dark matter particles) in arcmin, $\theta$ is the angle in degrees between the major axis and the projection of the galaxy in the simulation $x-y$ plane, $e$ is ellipticity defined as one minus the ratio between major and minor axis length, $r_h$ is the half-density/mass radius, and $N_{\star}$ is the number of stars the fitting procedure assigned as galaxy members. The true number of stars is indicated in the top left corner of each panel.}
    \label{fig:profile_23mag400kpc}
\end{figure*}

Figure~\ref{fig:profile_23mag400kpc} shows a typical example of one fitting procedure for the DES magnitude limit (r=23) at $400\,{\rm kpc}$ used throughout the main text. Note that this is only one out of a 100 fits used to determine the distributions presented in Figure~\ref{fig:rhalf_FeH_MV}.
The figure include the distribution of stars (top panel, galaxy members in red points), the observed stellar number density (black points including Poisson noise), the resulting profile (red line), and parameters with errors from the fitting procedure (bottom panel).

\begin{figure*}
    \centering
    \includegraphics[width=\linewidth]{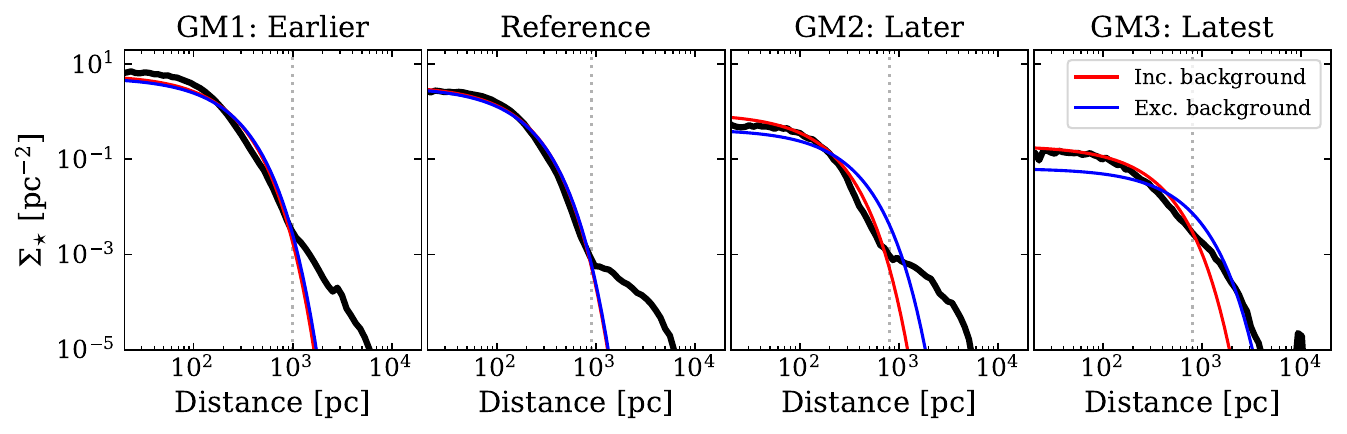}
    \caption{The projected surface number density of stars in our simulations. The gray dashed lines denote where the profile visually deviates from an exponential (extended halo stars). Red and blue lines show the exponential profile (Equation~\ref{eq:exp_profile} without the second term) using parameters obtained from the fitting procedure (Appendix~\ref{sec:LuminosityAndStructure}) with and without a population of background stars (see text for details), respectively.}
    \label{fig:profile_nobg}
\end{figure*}

Figure~\ref{fig:profile_nobg} shows the surface number density of each simulation calculated using spherical radii. The gray dashed lines denote the radius where we find a transition into an extended diffuse halo through visual inspection, highlighting multiple structural components in our galaxies. This transition results from galaxy growth in a cosmological environment (in particular through mergers) and makes fitting the profile in Equation~\ref{eq:exp_profile} challenging. This extended halo can not be distinguished when we artificially add a sample of background stars, but affects the parameters derived through the fitting method.

\begin{figure*}
    \centering
    \includegraphics[width=\linewidth]{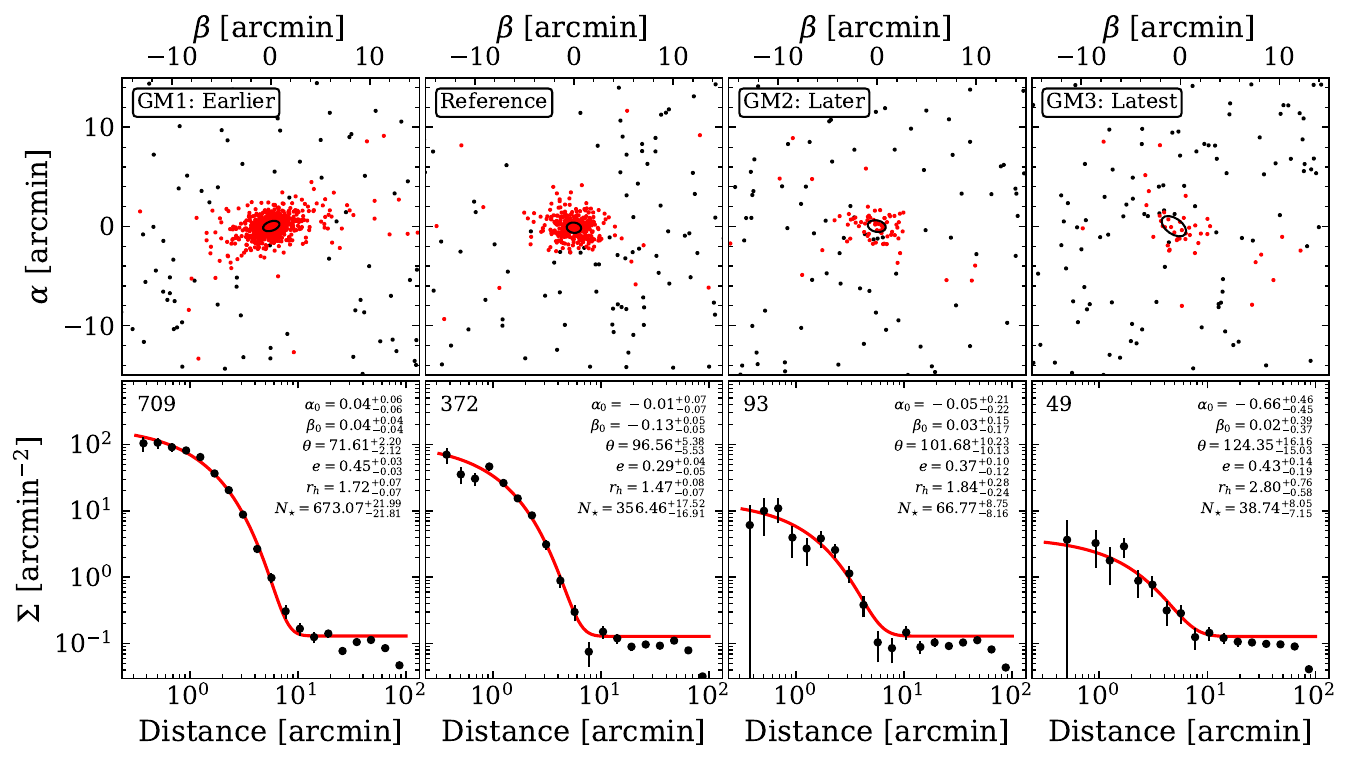}
    \caption{Same as Figure~\ref{fig:profile_23mag400kpc} but for 22 magnitudes in r-band at $200\,{\rm kpc}$}
    \label{fig:profile_22mag200kpc}
\end{figure*}

\begin{figure*}
    \centering
    \includegraphics[width=\linewidth]{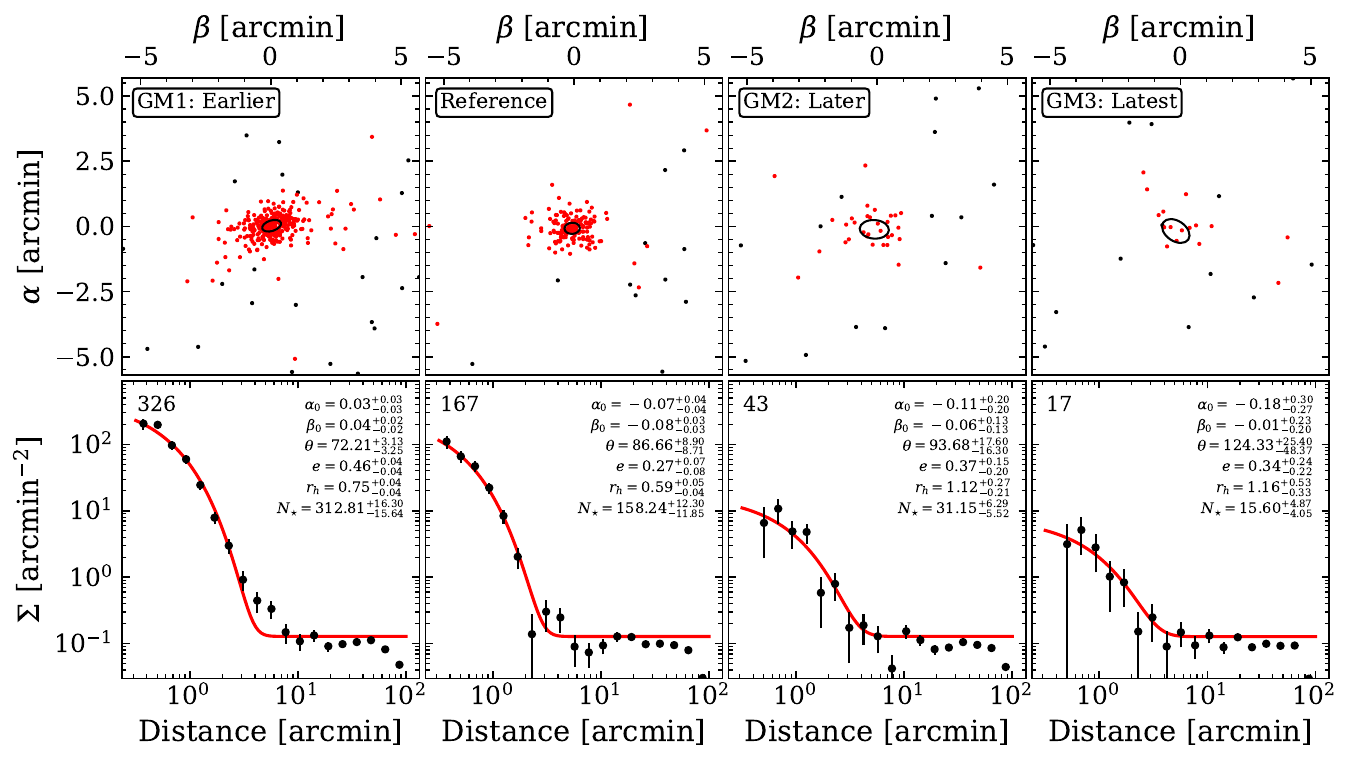}
    \caption{Same as Figure~\ref{fig:profile_22mag200kpc} but for 24 magnitudes in r-band at $1000\,{\rm kpc}$}
    \label{fig:profile_24mag1000kpc}
\end{figure*}

Throughout the main text, we presented our results as if observed at $400\,{\rm kpc}$ with a limiting magnitude in r-band of 23. In this way, our estimates suffer from the same limitations as if observed with state-of-the-art facilities and therefore directly comparable with the data shown by error bars in Figure~\ref{fig:rhalf_FeH_MV} \citep{Simon2019}. In addition, we performed the fitting procedure using SDSS limits (r-band limit of $22$ mag, Figure~\ref{fig:profile_22mag200kpc}) and the upcoming LSST (single pointing with r-band limit of $24$ mag, Figure~\ref{fig:profile_24mag1000kpc}). Similarly to the main text, we place the galaxies at a distance of $200\,{\rm kpc}$ ($1\,{\rm Mpc}$) where the smallest galaxy is barely observable with SDSS (LSST) to provide constraints for a range of observational quality.

\begin{figure*}
    \centering
    \includegraphics[width=\linewidth]{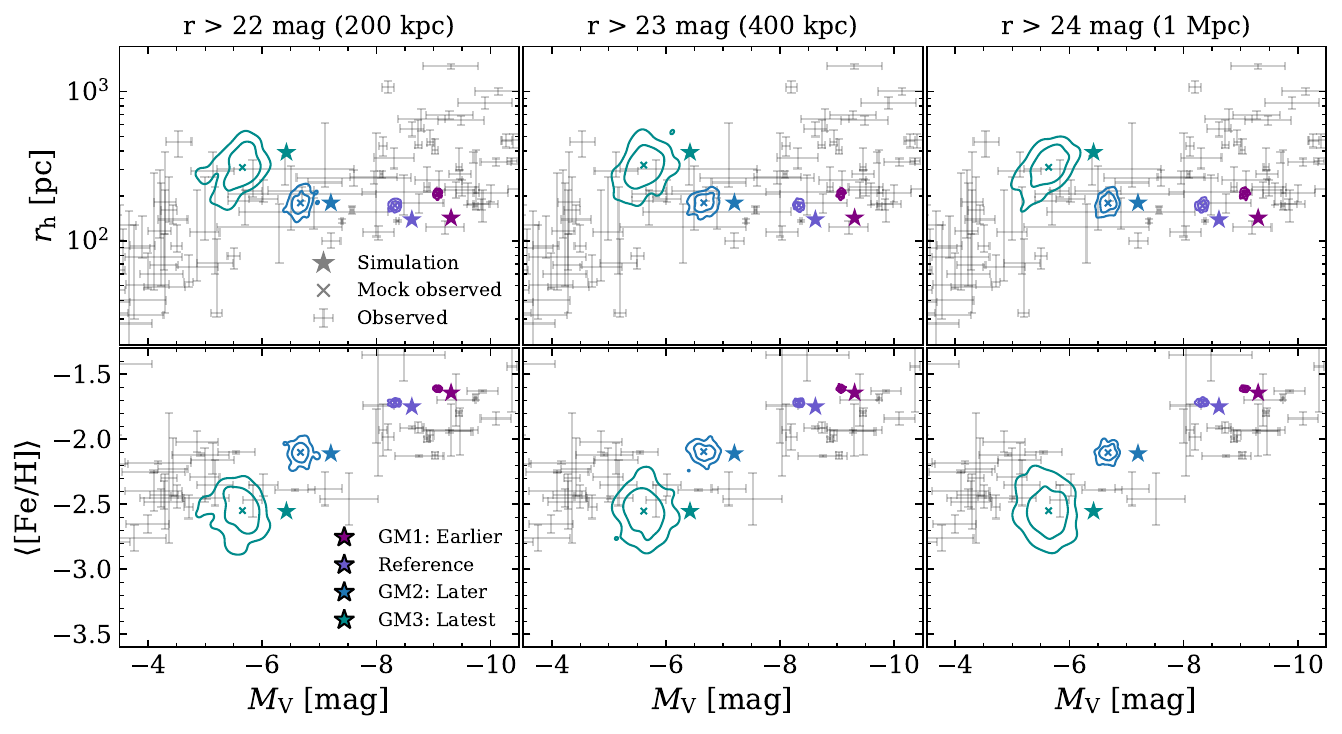}
    \caption{Each column is same as Figure~\ref{fig:rhalf_FeH_MV}, showcasing r-band limiting magnitudes at 22 ($200\,{\rm kpc}$), 23 ($400\,{\rm kpc}$), and 24 ($1\,{\rm Mpc}$), i.e., typical point-source magnitude limits for SDSS, DES, and the LSST at distances where the smallest galaxy (GM3: Latest) is at the limit of observability. Note that the center column is identical to Figure~\ref{fig:rhalf_FeH_MV}.}
    \label{fig:rhalf_FeH_MV_panels}
\end{figure*}

For completeness, Figure~\ref{fig:rhalf_FeH_MV_panels} presents the half-light radius and average [Fe/H] as functions of total V-band magnitude for all three facilities. Our conclusions remain the same for all three cases, only showing an increase in the scatter in the estimated observables directly related to the number of observed stars.

\bibliography{ref}{}
\bibliographystyle{aasjournal}

\end{document}